\documentclass[a4paper,11pt]{article}
\usepackage{pos}
\usepackage{cleveref}

\newcommand{\lt}{\tau_{\rm L}} 
\newcommand{\rt}{x_0}    
\newcommand{\f}{\phi}    

\title{Kernel controlled real-time Complex Langevin simulation}

\author*{Daniel Alvestad}
\author{Rasmus Larsen}
\author{Alexander Rothkopf}

\affiliation{University of Stavanger}

\emailAdd{daniel.alvestad@uis.no}

\abstract{This study explores the utility of a kernel in complex Langevin simulations of quantum real-time dynamics on the Schwinger-Keldysh contour. We give several examples where we use a systematic scheme to find kernels that restore correct convergence of complex Langevin. The schemes combine prior information we know about the system and the correctness of convergence of complex Langevin to construct a kernel. This allows us to simulate up to $1.5\beta$ on the real-time Schwinger-Keldysh contour with the 0+1 dimensional anharmonic oscillator using $m=1$, $\lambda=24$, which was previously unattainable using the complex Langevin equation.}

\FullConference{%
The 39th International Symposium on Lattice Field Theory,\\
8th-13th August, 2022,\\
Rheinische Friedrich-Wilhelms-Universität Bonn, Bonn, Germany
}

\begin{document}
\maketitle

\section{Introduction}

Understanding the real-time dynamics of strongly correlated quantum systems is of big interest to many disciplines both in high and low-energy physics. The underlying problem is the sign-problem, which has been proven to be an NP-hard problem, meaning that we need a system-specific solution to overcome it.

One of the methods to remedy the problem is the complex Langevin method where the fields are complexified and then evolved in a fictitious time by a stochastic differential equation. The method reproduces, for a limited set of parameter ranges, correct results in several systems with complex actions. One of them is the simulation of QCD at finite chemical potential\cite{Attanasio:2022mjd}. The method has suffered from two major drawbacks, the occurrence of unstable trajectories (runaway solutions) and convergence to the wrong solution. In strongly correlated system it was shown that the first one can be avoided by the use of implicit solvers\cite{Alvestad:2021hsi}. The latter is still an unsolved problem, and this study will propose a novel method to correct the convergence based on finding optimal kernels in the complex Langevin equation based on prior information. 

Kernels have been studied to remedy the convergence problem of the complex Langevin in \cite{Okamoto:1988ru,Okano:1991tz} for simple systems. These studies seem promising for the simple system studied, but extending the use of kernels to more complicated models was difficult. We propose a strategy to optimize kernels based on the use of prior knowledge of the model and the correctness criterion of the complex Langevin. For a full overview of the method, we refer to \cite{alvestad2022, alvestad2022_conf}, which explains the method in detail. In this paper, we will first give an overview of the strategy for the optimized kernels, then discuss the kernels in relation to the correctness criterion and the Lefschetz thimbles. 

\section{Kernelled complex Langevin}

The general framework for the complex Langevin equation (CLE) is based on Stochastic quantization, where we construct a stochastic process evolving the fields in a fictitious time $\tau_{\rm L}$. The stochastic differential equation is dependent on a drift term, $dS/d\phi$ and a noise structure such that it correctly reproduces the fluctuations of the original theory. In terms of the path-integral, the expectation values of the observables of the system 
are given by
\begin{align}\label{eq:PathIntegralM}
    \langle O \rangle = \frac{1}{Z} \int {\cal D}\phi\;O[\phi] e^{iS_M[\phi]}, \quad S_M[\phi]=\int d^dx L_M[\phi],
\end{align}
where we have noted the Minkowski time action as $S_M$. The stochastic process, called the complex Langevin equation, which is deployed for this path integral is given by
\begin{equation}\label{eq:CLE}
\begin{aligned}
    & \frac{d\f}{d\lt} = i\frac{\delta S_M[\f]}{\delta \f(x)} + \eta(x,\lt) \quad \textrm{with}   \\
    &\langle \eta(x,\lt) \rangle = 0, \quad \langle \eta(x,\lt) \eta(x',\lt') \rangle = 2\delta(x-x')\delta(\lt-\lt').
\end{aligned}
\end{equation}
Here the fields are complexified $\phi \rightarrow \phi_R + i\phi_I$. 

After complexifying the fields, the question of correct convergence needs to be answered. Two main criteria need to be satisfied in order for the CL to converge to the correct solution; these are summarised in the correctness criterion \cite{Aarts:2011ax}. The first one is that the late time complex distribution of the field $\Phi[\phi]$ needs to converge to $\Phi \propto \exp\left( iS_M \right)$ to reproduce the model, and the second one is that there should be no boundary terms \cite{Scherzer:2018hid}. We will discuss the criteria in more detail in \cref{sec:ThimblesCorrectness}. 


We know from the Fokker-Planck equation 
of the real Langevin equation that there exists freedom in the definition of the equation such that it still converges to the original theory, i.e, the late time distribution is still $\exp\left(-S_E\right)$ (Here we have used the Euclidean formulation of the action $S_E$ to describe a real action). In the real case, introducing a kernel changes the approach to the unique stationary distribution, while not changing its form.

On the other hand, when we introduce a complex kernel in the complex Langevin equation, we obtain a non-neutral modification of the dynamics, meaning that we can also change the stationary distribution of the dynamics. In this paper, we will restrict the kernel to be field-independent, i.e., it is a constant matrix in Langevin time, mixing the field and noise at different space and time coordinates in the Fokker-Planck equation. The corresponding complex Langevin equation is then given by
\begin{equation}
    \frac{\partial \f(x)}{\partial \lt} = K(x)\frac{\partial S[\f]}{\partial \f(x)} + H(x)\eta(x,\lt), \quad \langle \eta(x,\lt)\eta(y,\lt') \rangle = 2\delta(x-y)\delta(\lt - \lt')
\end{equation}
where we have used the kernel $K(x)$ and its factorization $H\;H^T=K$. Note that some studies absorb the quantity $H$ into the noise term $\eta(x,\lt)$. 

\section{Learning optimized kernel}
\label{sec:learnOptmizedKernel}
We will, in this section, introduce a strategy on how to construct kernels systematically in such a way as to improve the convergence of the real-time complex Langevin. We will use prior knowledge available, such as the symmetries of the system, the Euclidean correlator accessible in conventional simulations, and the correctness criterion. We set up a cost function, combining each type of prior information, and minimizing it by updating the kernel's parameters. In this study we use a combination of the symmetries and prior known Euclidean correlators, such that our cost function is $L^{\textrm{Prior}} = L^{\textrm{sym}} + L^{\textrm{BT}}$.

To minimize the loss by updating the kernel parameters, we need to evaluate the gradient of the loss function with respect to the kernel parameters. This will involve taking the gradient of the field with respect to every kernel parameter throughout the whole CL simulation. This is, in principle, possible using auto-differentiation (AD) and can be implemented for one-degree-of-freedom models with a small number of kernel parameters. The cost of direct AD methods however grows quickly with system size and  a possible way forward is to implement an adjoint sensitivity SDE method specifically for the real-time problem, which is beyond the scope of this study. 

Instead, we construct a loss function that we can easily calculate an approximate gradient of. The loss function is inspired by the successful dynamic stabilization approach \cite{attanasio_dynamical_2019}, where one guides the drift term towards the non-complexified part of the action. The loss function we use is 
\begin{equation}\label{eq:driftLoss}
    L_{D} =  \left\langle \left|  \frac{1}{N}\sum_i^N D(\f_i) \cdot (-\f_i) - |D(\f_i)||\f| \right| \right\rangle 
\end{equation}
where $D=K\partial S / \partial \phi$ which pulls the field degrees towards the origin. By minimizing this function, we will decrease the boundary terms and attempt to guide the kernelled complex Langevin towards achieving correct convergence. We evaluate the gradient of $L_D$ by treating every configuration as independent events, and hence do not take the gradient of the whole CL simulation. This is an approximation to the true gradient, as the field depends on the rest of the simulation. However, we find that in practice it works well to guiding the kernel parameters in the correct direction. 

We now apply the low-cost gradient in an iterative optimization scheme\footnote{In this paper we use the ADAM optimizer}, where we in every update step monitor the $L^{\textrm{prior}}$ loss function to see that we are progressing in the right direction. We select in the end the kernel with the smallest $L^{\textrm{prior}}$.

\subsection{Anharmonic oscillator case study}

In this section, we will use the strategy laid out above to construct a kernel for the anharmonic oscillator in $0+1D$ on the thermal Schwinger-Keldysh contour at short to intermediate real-times. We extend correct convergence to real-times where previously no appropriate kernel has been found. The model has been studied by complex Langevin in \cite{berges_real-time_2008, Alvestad:2021hsi} where it was shown that there is a limit to the real-time extent before the naive complex Langevin converges to the wrong solution. The action of the model is given by
\begin{equation}
    \label{eq:AHOAction}
    S(\f) = \int d\rt \left\{ \frac{1}{2}  \left( \frac{\partial \f(\rt)}{\partial \rt} \right)^2 - \frac{1}{2}m^2\f^2(\rt) - \frac{\lambda}{4!}\f^4(\rt) \right\}
\end{equation}
where we use $\lambda=24$ in this paper. The real-time contour deployed is the canonical version first used for the complex Langevin in \cite{Alvestad:2021hsi}. This contour follows the real axis in the forward and backward part of the contour, i.e., no tilt is applied for regularizing the theory. For the Euclidean part, we follow the imaginary axis down to $-i\beta$, which is set to $\beta=1$ in this paper. We discretize the contour with 20 time-points for every 1 in time, such that a contour with a maximum real-time of $1$ will have 20 points in the forward part, 20 points in the backward part, and then 20 points along the Euclidean part, yielding a total of 60 points along the full contour. 

The field-independent kernel is parameterized by a $\tau_L$ independent complex matrix $K$ with $(2N_{\rm RT}+N_{\rm E})^2$ entries, multiplying the $2N_{\rm RT}$ d.o.f. on the forward and backward contour and the $N_{\rm E}$ ones on the imaginary time branch. We tune these kernel parameters when running the optimization scheme to achieve optimal convergence. 

\begin{figure}\centering
    \includegraphics[scale=0.35]{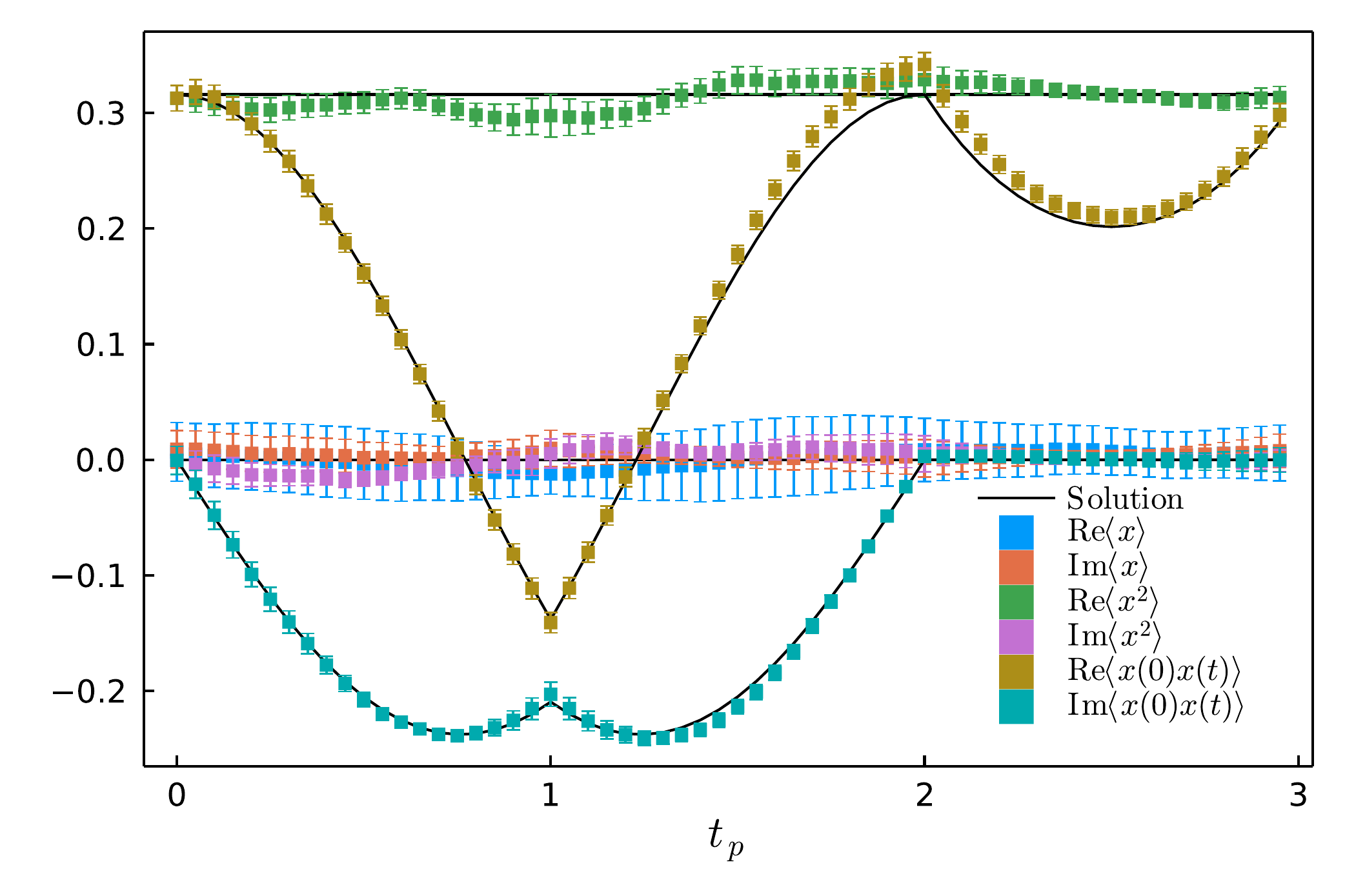}
    \includegraphics[scale=0.35]{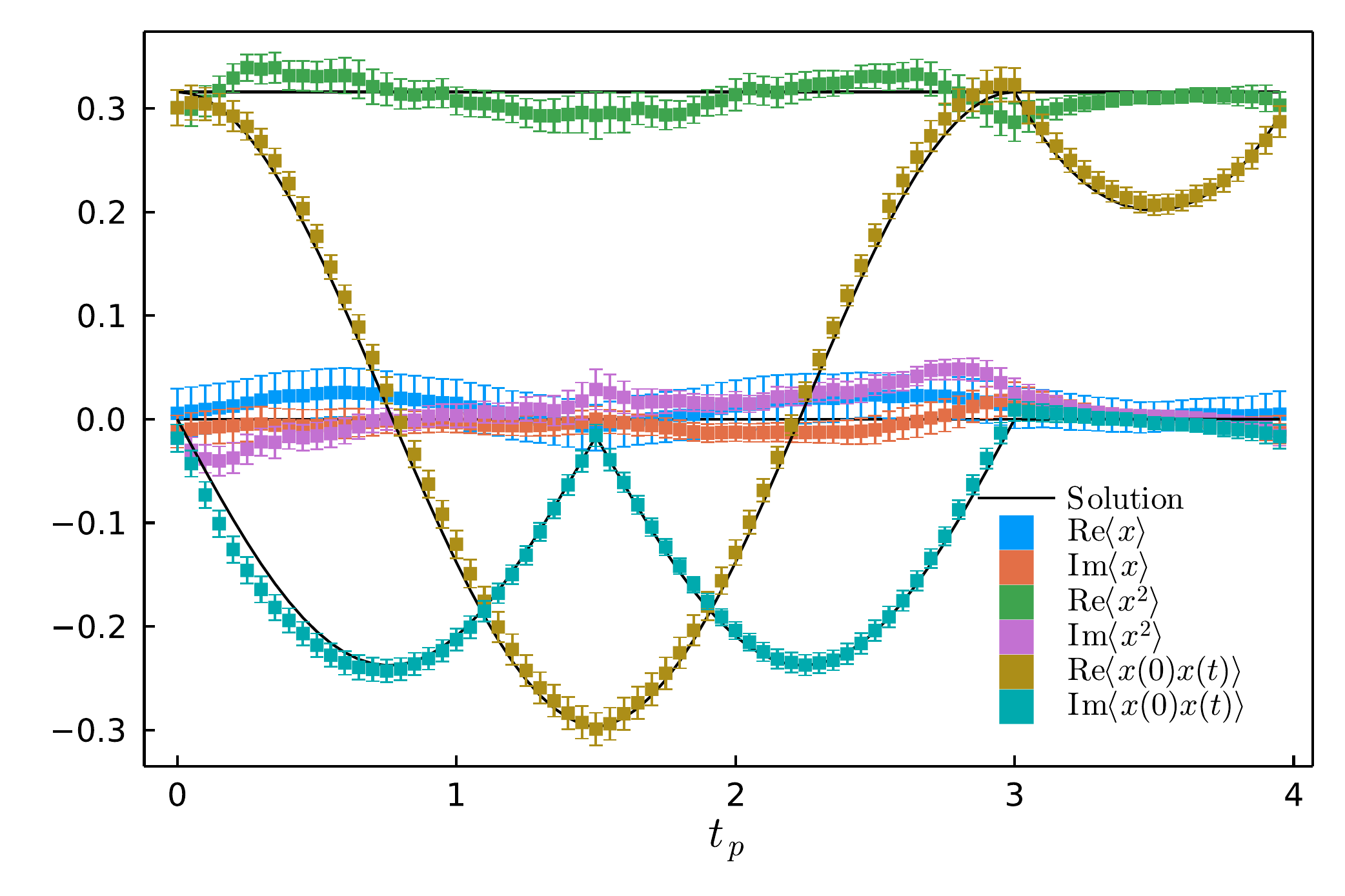}
    \caption{Two different real-time extents; $mt^{\textrm{max}}=1.0$ (left) and $mt^{\textrm{max}}=1.5$ (right) showing the CLE simulation for the optimized kernel. We show three observables $\langle x \rangle$, $\langle x^2 \rangle$ and the correlator $\langle x(0) x(t) \rangle$ plotted against the contour parameter $t_p$, which for $mt^{\textrm{max}}=1.0$ is $t_p = 1$ at $mt =1.0$ and then $t_p=2.0$ at $mt=0.0$ after the backward path of the contour.}
    \label{fig:ResultObsIntPlots}
\end{figure}

The kernel optimization scheme is tested for the anharmonic oscillator on a Schwinger-Keldysh contour with two different real-time extents. The results can be seen in \cref{fig:ResultObsIntPlots}, wherein the left panel we show $mt^{ \textrm{max} } = 1.0$ and the right panel show $mt^{ \textrm{max} } = 1.5$. We see that by optimizing the kernel based on the low-cost gradient and monitoring based on $L^{\textrm{Prior}}$ we manage to extend the correct convergence of CL up to $mt^{\textrm{max}}=1.5$.  

There are, however, limitations to the use of the low-cost gradient approximation, as using the same scheme for a longer real-time extent of $mt^{\textrm{max}}=2.0$ fails to restore the convergence of CL \cite{alvestad2022}. One reason for this failure is that for longer real-time extents, minimizing $L_D$, i.e., pulling the drift towards the origin, is minimizing the boundary terms but introducing a new stationary distribution other than $\exp(iS_M)$. This breaks the correctness criterion. We will closely examine the relationship between the kernel and correctness criterion, and give more detail on why only minimizing the boundary terms is not sufficient in the following section.

\section{Thimbles, boundary condition and kernel}\label{sec:ThimblesCorrectness}
In this section, we will be investigating one-degree-of-freedom models, for which, in the literature, optimal kernels are known. We use the action $S=\frac{1}{2}\sigma x^2 + \frac{\lambda}{4}x^4$, which leads to the following partition function $Z = \int d\f e^{-S}$, which was studied in \cite{Okamoto:1988ru,Okano:1991tz,Aarts:2013fpa}. The model is interesting since it exhibits similar properties as the real-time anharmonic oscillator; the convergence problem appears, breaking both the boundary term condition and the equilibrium distribution of the Fokker-Planck equation for various parameters. In this section, we would like to understand better how kernel affects the behavior of the complex Langevin, and also how this relates to the Lefschetz thimbles and the correctness criterion \cite{Aarts:2011ax}.

Let's first take the simplest example where $\sigma = i$ and $\lambda=0$, which was shown in \cite{Okamoto:1988ru} to have the optimal kernel of $K=-i$. By applying this kernel to CLE it turns the drift into the simple form of $K\frac{\partial S[x]}{\partial x}=-x$ and we end up with a complex noise given by the coefficient $H = \sqrt{-i} = e^{-i\frac{\pi}{4}}$. This correspond to the complex Langevin equation sampling from a straight line $z(x)=xe^{-i\frac{\pi}{4}}$ which happens to be exactly the same as the Lefschetz thimble. We obtain the thimble by solving the thimble flow equation analytically \cite{Woodward:2022pet}
\begin{equation}\label{eq:ThimbleFlow}
    \frac{d x}{d\tau} = \overline{\frac{d S[x]}{d x}}.
\end{equation}
This points to a connection between the flow of the thimbles and the kernel. The connection is, however, not as trivial for non-linear actions which we will discuss next. 

We now take a closer look at two specific sets of parameters. The first one is $\sigma=4i$ and $\lambda = 2$ where we can find an optimal kernel, and as the second one, we choose $\sigma=-1 + 4i$ with the same $\lambda=2$, were for correct convergence, we have to go beyond a constant, field-independent kernel. 

The first parameter set was found in \cite{Okamoto:1988ru} to require an optimal kernel of $K_1=\exp\{-i\frac{\pi}{3}\}$ to regain correct convergence. However, if we use the optimization scheme laid out in \cref{sec:learnOptmizedKernel} and minimize using the approximate loss function $L_D$, we find two minima; $K_1=\exp\{-i\frac{\pi}{3}\}$ and $K_2 = \exp\{-i\frac{2\pi}{3}\}$. We have parameterized the kernel by a phase rotation, leaving only one parameter $\theta$ to be tuned, and the kernel form is $K=e^{i\theta}$. The second minimum does not give correct convergence, and we will now see if we can distinguish the correct convergence kernel $K_1$ from $K_2$ using the correctness criterion \cite{Aarts:2011ax}. In \cite{alvestad2022}, we give an overview of the correctness criterion when using a kernel. 

\begin{figure}\centering
    \includegraphics[scale=0.04]{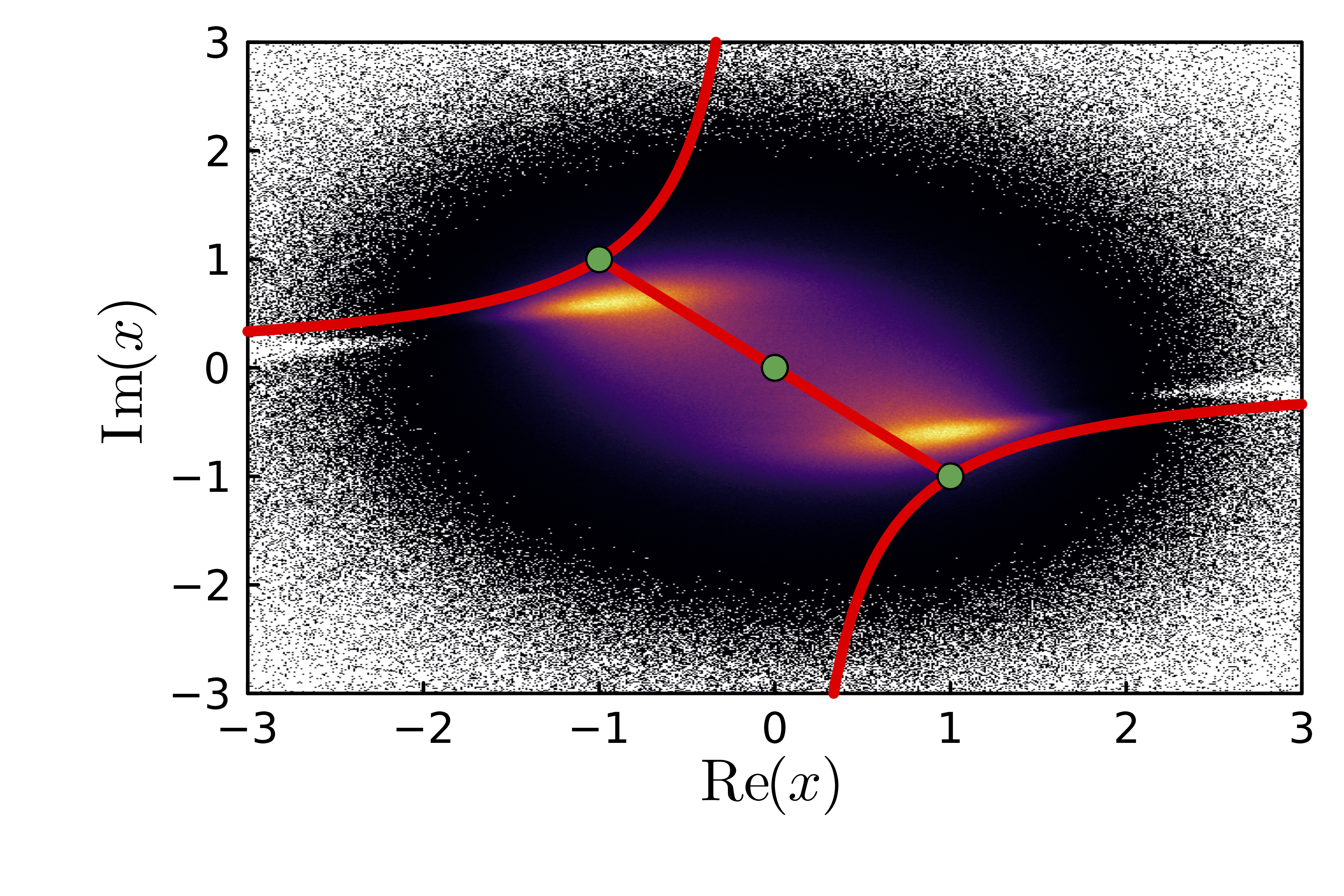}
    \includegraphics[scale=0.04]{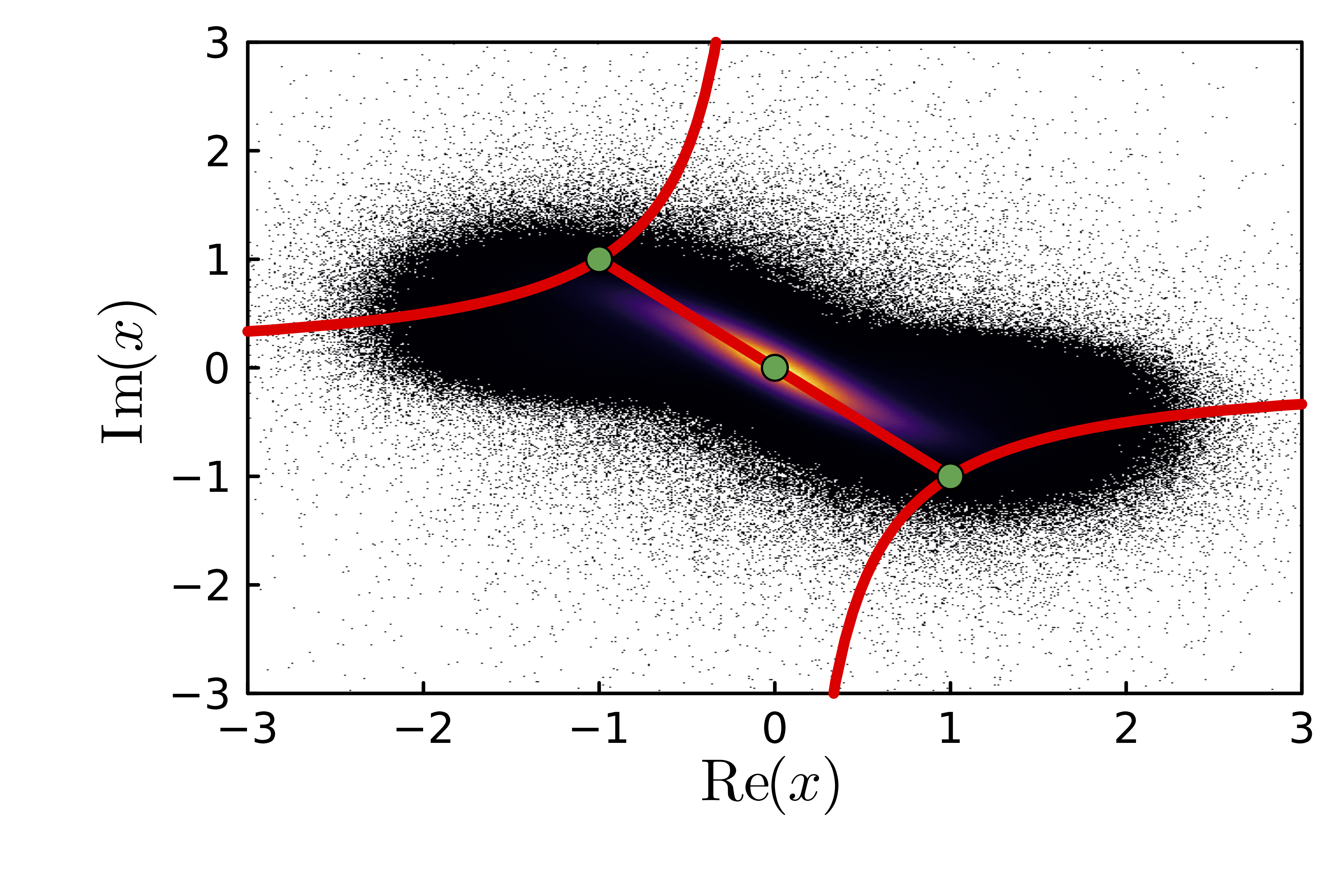}
    \includegraphics[scale=0.04]{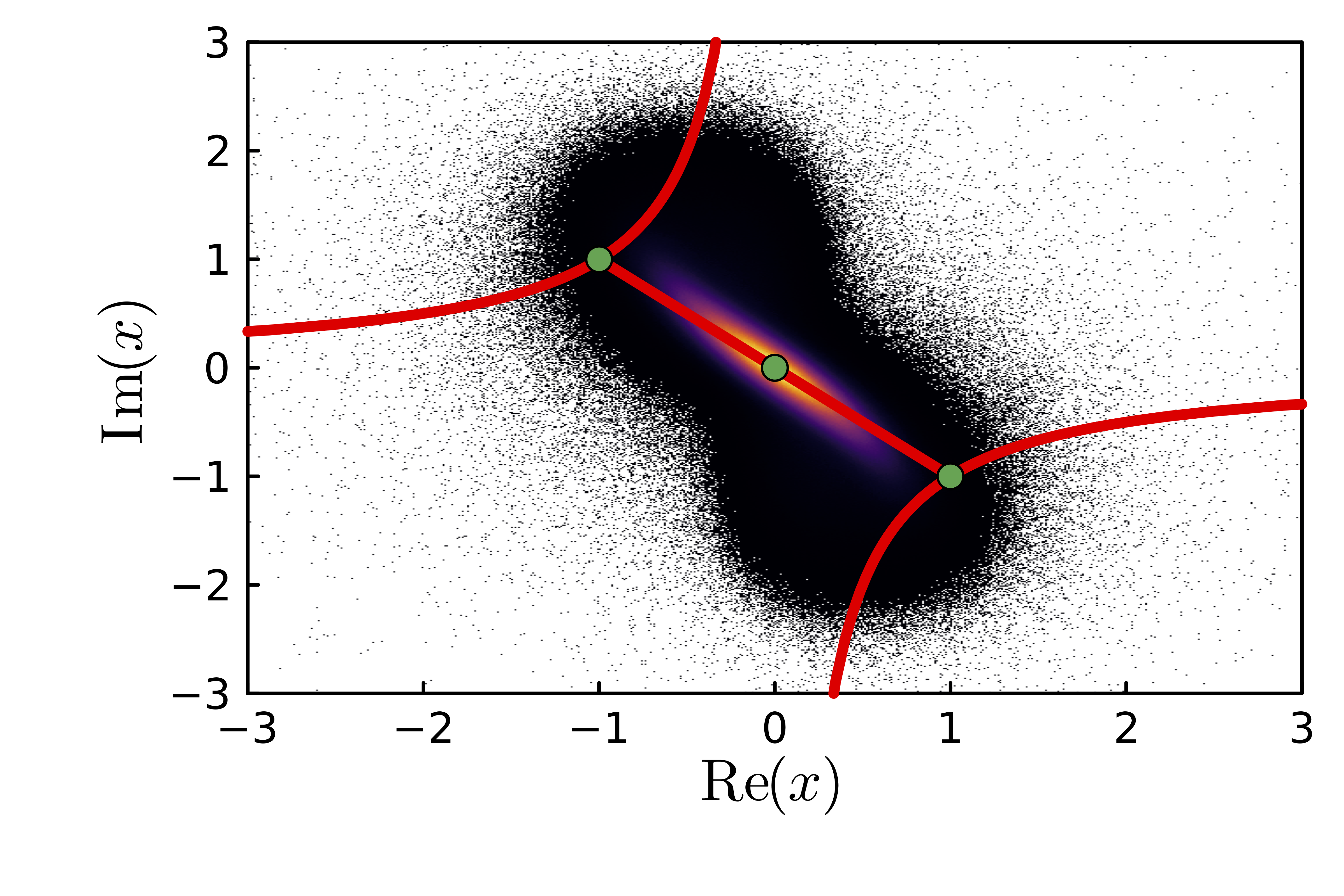}
    \includegraphics[scale=0.28]{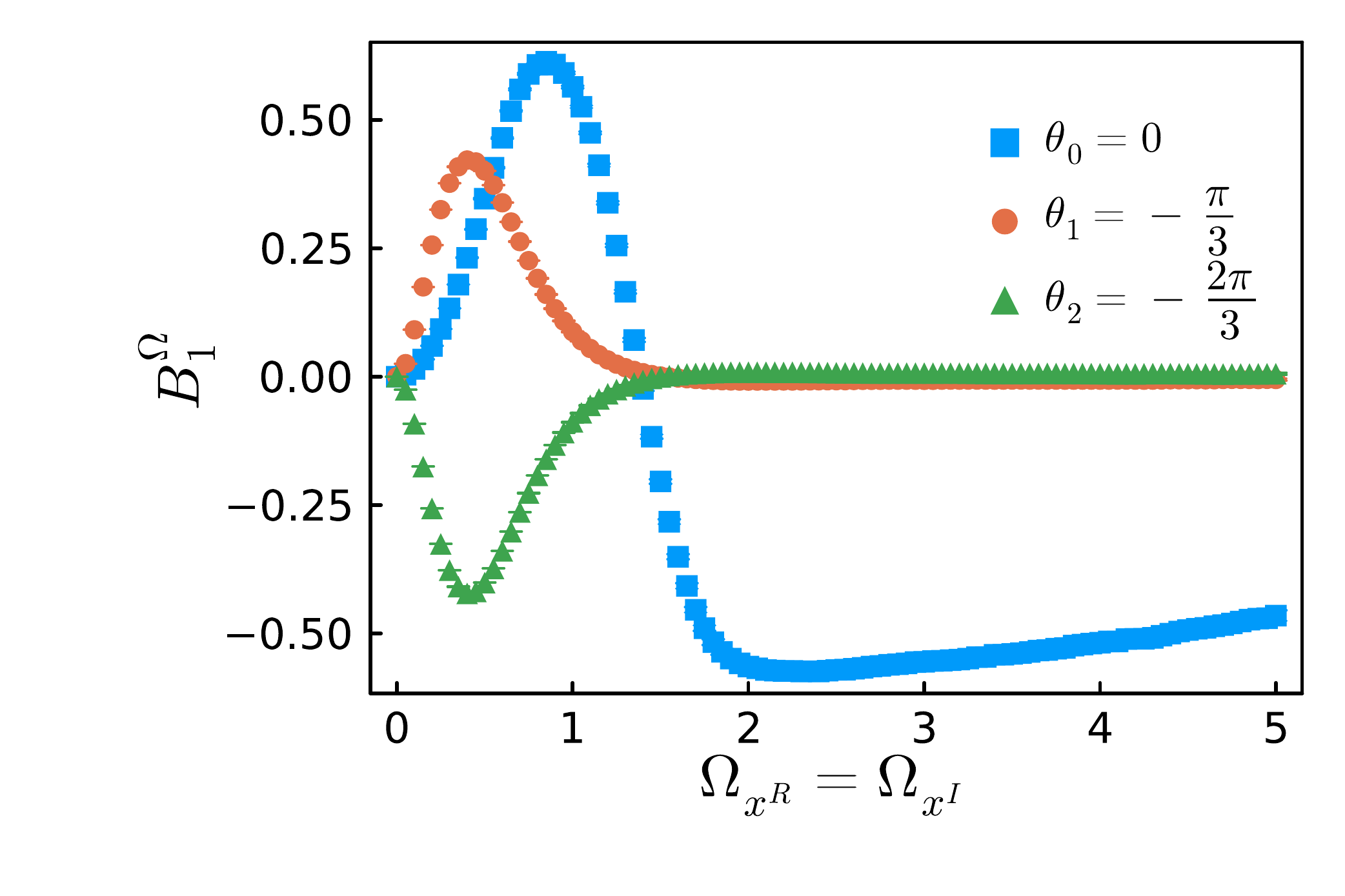}
    \caption{Distribution of the complex Langevin simulation and the Lefschetz thimble (red line) with $\sigma =4i$ and $\lambda=2$, using different kernels; $K_0=1$ (top left), $K_1={\rm exp}[-i\pi/3]$ (top right) and $K_2={\rm exp}[-i2\pi/3]$ (bottom left). The green points denote the critical points. The color in the distribution heat map corresponds to the number of samples at the corresponding position (a lighter color refers to a higher value). In the lower right pane, we show the boundary terms value for the three different kernels.}
    \label{fig:Dist042}
\end{figure}

We show the complex Langevin distribution together with the corresponding Lefschetz thimbles in \cref{fig:Dist042} for the three different kernels; $K_0=1$(top left), $K_1$(top right) and $K_2$(bottom left). We see that the distribution of the identity kernel $K_0$ is broad and that the peaks of the distribution point in the direction of the real axis. This is in line with the findings from the previous simple model where the angle of the noise coefficient, which in this case is along the real axis, will be the dominant direction of the CLE sampling. We see a more localized distribution for the two other kernels, which is rotated in the same direction as the thimble. The main difference between the two is that $K_1$ bends towards the part of the thimble going towards the real axis, while $K_2$ bends towards the thimble going towards the imaginary axis. 

We then plot the boundary terms\cite{Scherzer:2018hid} for the real part of the $\langle x^2 \rangle$ observable in the lower right plot of \cref{fig:Dist042}. See \cite{alvestad2022} for a derivation of the boundary terms with a kernel. Here we clearly see that the identity kernel $K_0$ has boundary terms appearing, while for $K_1$ and $K_2$ there are no boundary terms. This means we have found a case where we have no boundary terms, and CLE still converges to the wrong solution. The failures of convergence for $K_2$ are therefore assigned to it violating the criterion that the equilibrium distribution of the Fokker-Planck equation should be $\exp\{-S\}$\cite{alvestad2022}. 
 
\begin{figure}\centering
    \includegraphics[scale=0.04]{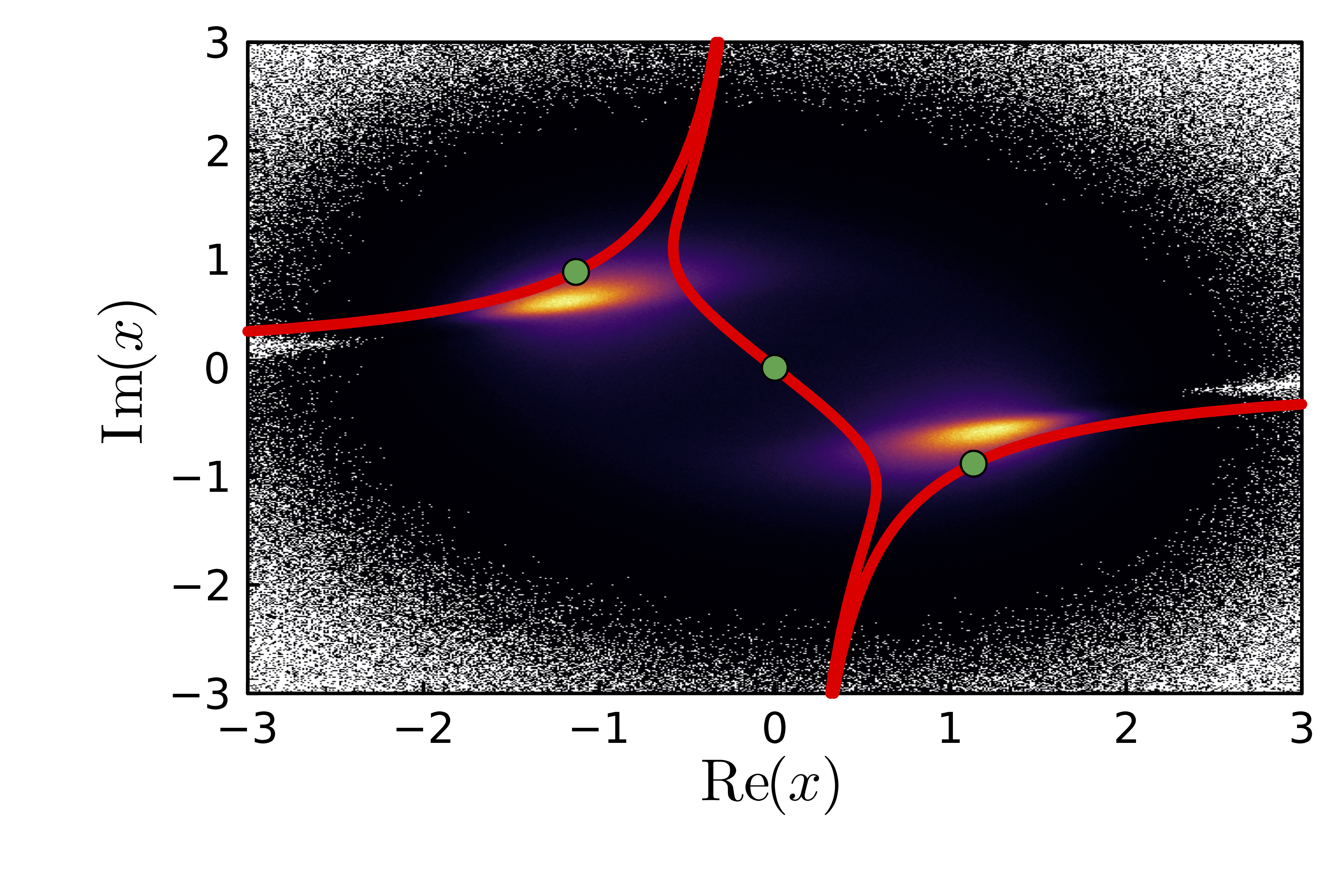}
    \includegraphics[scale=0.04]{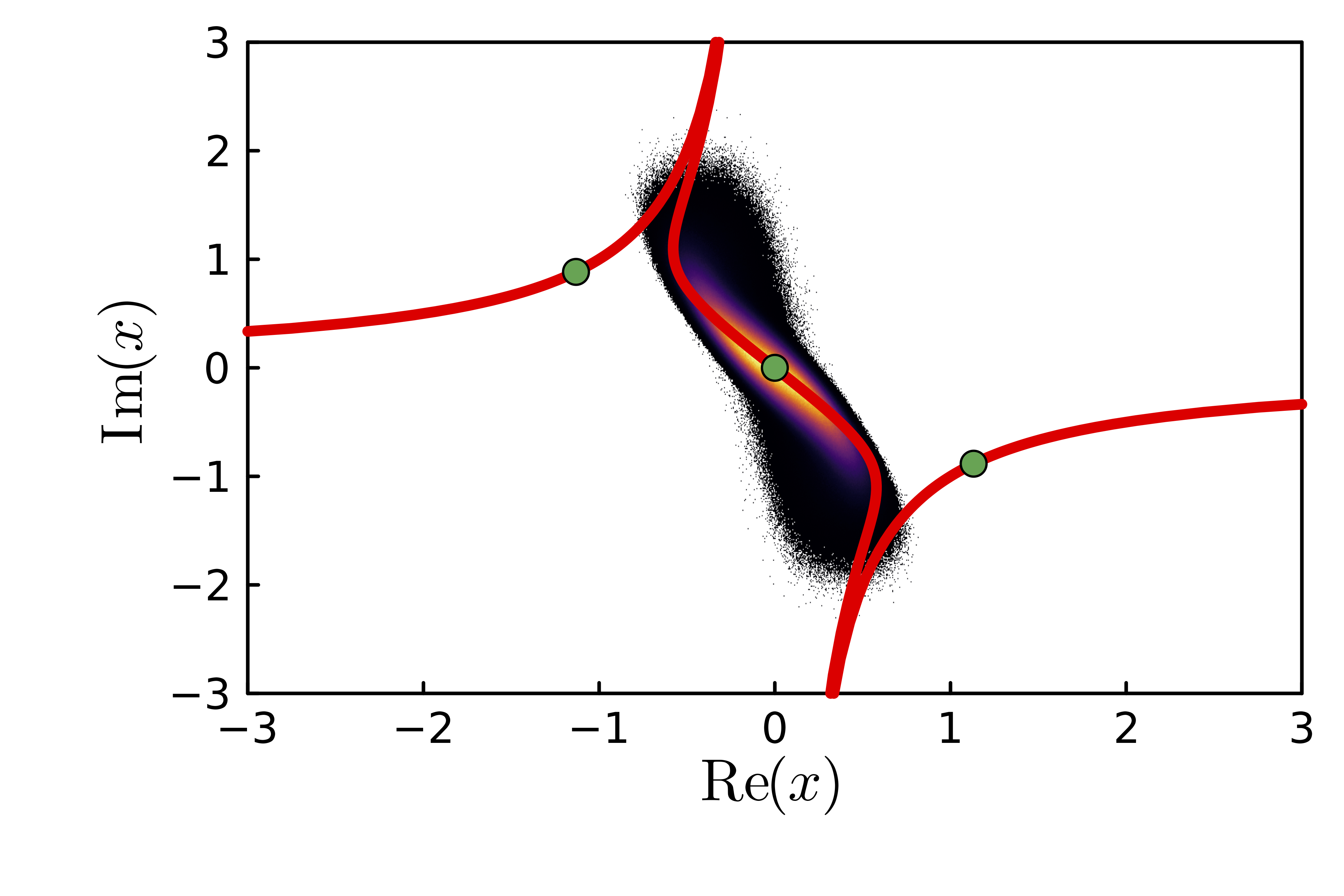}
    \includegraphics[scale=0.04]{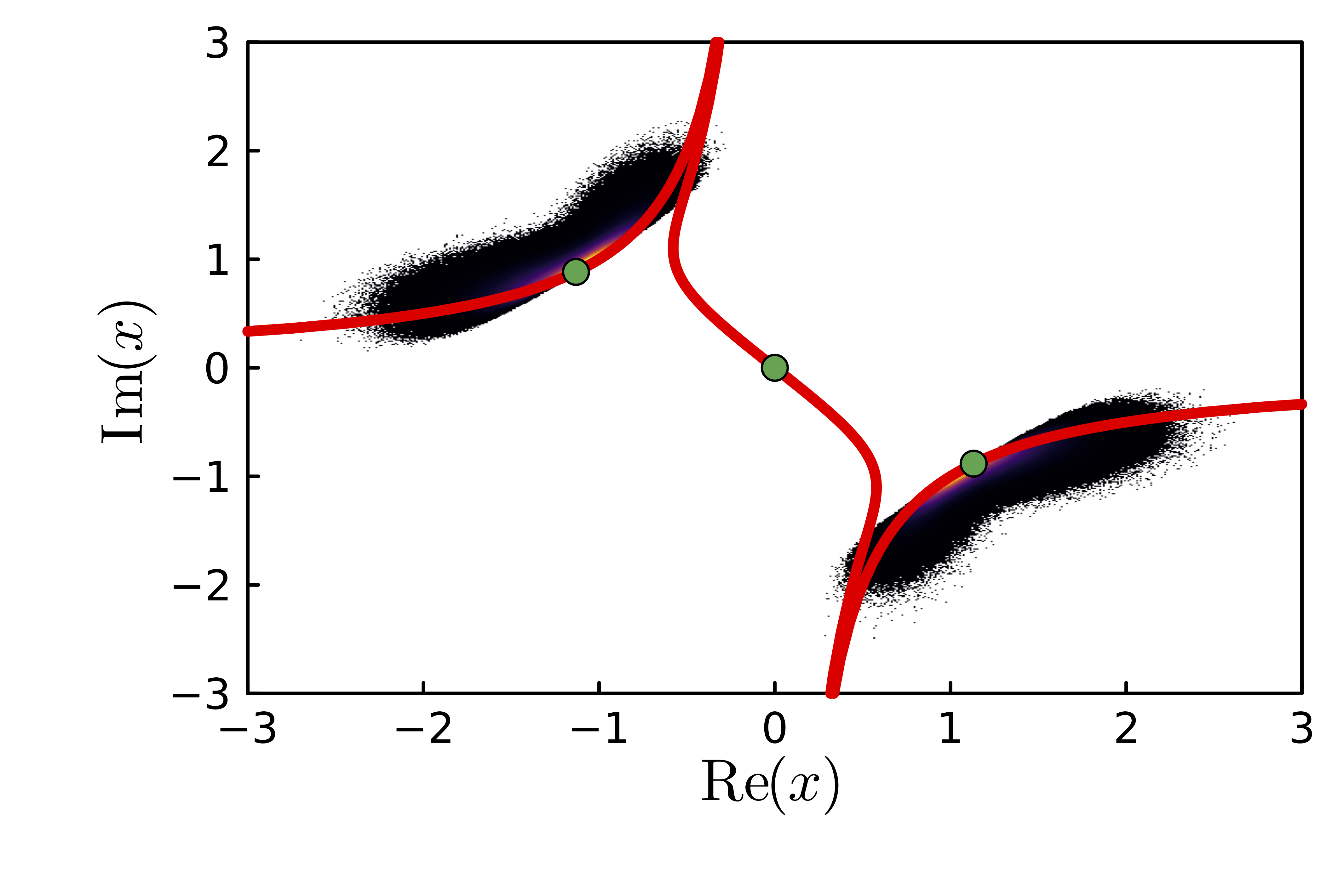}
    \includegraphics[scale=0.28]{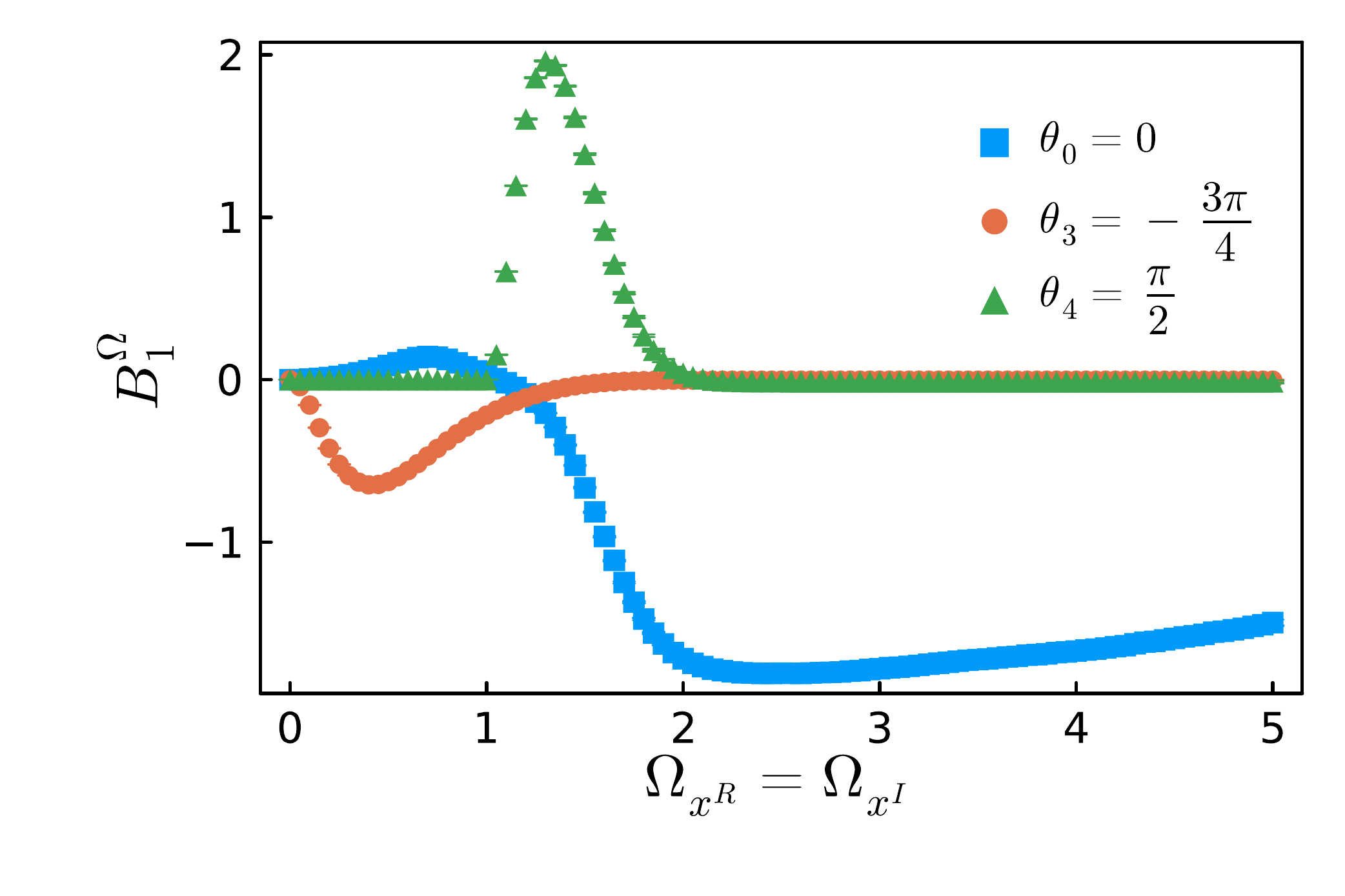}
    \caption{Distribution of the complex Langevin simulation and the Lefschetz thimbles (red line) with $\sigma = -1 + 4i$ and $\lambda=2$, using different kernels; $K_0=1$ (top left), $K_3=e^{-i\frac{3\pi}{4}}$ (top right) and $K_4=e^{i\frac{\pi}{2}}$ (bottom left). In the lower right pane we show the boundary terms for the three different kernels.}
    \label{fig:Dist-142}
\end{figure}

For the second parameter point $\sigma = -1+4i$ and $\lambda = 2$, we have a similar effect when minimizing for $L_D$. In this case, there are more than two minima, but we have selected two interesting minima, both of which have no boundary terms while not converging correctly. We have selected these because they yield interesting behavior in the distributions compared to the flowed thimbles, which can be seen in \cref{fig:Dist-142}. The kernels have the parameters $\theta_3=\frac{-3\pi}{4}$ and $\theta_4=\frac{\pi}{2}$. The corresponding angle for these kernels is close to the direction of the thimble at the critical points (green points), which are given by the solution to $dS/dx = 0$. We see that the distributions of the two kernels are localized close to the critical points, for $K_3$, which is the critical point at the origin, and for $K_4$, the critical points that are away from the origin.

\section{Conclusion}
We have proposed a way to systematically optimize kernels based on prior information to restore the correct convergence of strongly correlated quantum systems on the real-time Schwinger-Keldysh contour. We have demonstrated the strategy by extending the extent of correct convergence beyond previous state-of-the-art simulations. Exploration of field-dependent kernels and implementing adjoint sensitivity methods for gradients of $L^{\textrm{prior}}$ on real-time complex Langevin are work in progress.

We have also investigated the connection between the application of a kernelled Langevin equation and Lefschetz thimbles. This showed that applying a kernel will change the CLE distribution based on the match between the kernel's angle and the thimble's angle at the critical points. We have also shown that there exist cases where it is not enough only to check that there are no boundary terms to make any statement of the correctness criterion being satisfied; we also need to check if the Fokker-Planck equilibrium distribution is $\exp\{iS_M\}$.  As the use of the Fokker-Planck equation in the evaluation of the correctness criterion is prohibitively expensive for system larger than one-degree of freedom, the use of other prior knowledge is important. 

\section*{Acknowledgements}
The team of authors gladly acknowledges support by the Research Council of Norway under the FRIPRO Young Research Talent grant 286883. The numerical simulations have been partially carried out on computing resources provided by  
UNINETT Sigma2 - the National Infrastructure for High Performance Computing and Data Storage in Norway under project NN9578K-QCDrtX "Real-time dynamics of nuclear matter under extreme conditions"  

\bibliographystyle{plain}
\bibliography{references}

\end{document}